\documentclass[review]{elsarticle}

\usepackage{moreverb}
\usepackage{verbatimbox}
\usepackage{array}
\usepackage[colorlinks,bookmarksopen,bookmarksnumbered]{hyperref}
\usepackage{multirow}
\usepackage[boxruled]{algorithm2e}
\usepackage{amsmath}
\usepackage{amssymb}
\usepackage{subcaption}
\usepackage{float}
\usepackage{graphicx}

\DeclareMathOperator*{\argmax}{arg\,max}

\newenvironment{megaalgorithm}[1][htb]
{% Update algorithm name
	\begin{algorithm}%
	}{\end{algorithm}}

\begin{document}

\begin{frontmatter}

\title{Improving SIEM capabilities through an enhanced probe for encrypted Skype traffic detection}

\author[address1]{Mario Di Mauro\corref{corresp}}
\author[address2]{Cesario Di Sarno}

\address[address1]{University of Salerno, Dept. of Engineering and Applied Maths, Fisciano (SA), Italy}
\address[address2]{Computer Science Research Group (COSIRE), Aversa, Italy}

\cortext[corresp]{Corresponding author}

\begin{abstract}
Nowadays, the Security Information and Event Management (SIEM) systems take on great relevance 
in handling security issues for critical infrastructures as Internet Service Providers.
Basically, a SIEM has two main functions: \emph{i)} the collection and the aggregation of log data and security information from disparate network devices (routers, firewalls, intrusion detection systems, ad hoc probes and others) and \emph{ii)} the analysis of the gathered data by implementing a set of correlation rules aimed at detecting potential suspicious events as the presence of encrypted real-time traffic.
In the present work, the authors propose an enhanced implementation of a SIEM where a particular focus is given to the detection of encrypted Skype traffic by using an ad-hoc developed enhanced probe (\emph{ESkyPRO}) conveniently governed by the SIEM itself.
Such enhanced probe, able to interact with an agent counterpart deployed into the SIEM platform, is designed by exploiting some machine learning concepts. 
The main purpose of the proposed ad-hoc SIEM is to correlate the information received by \emph{ESkyPRO} and other types of data obtained by an Intrusion Detection System (IDS) probe in order to make the encrypted Skype traffic detection as accurate as possible.
\end{abstract}


\end{frontmatter}

%\linenumbers

\section{Introduction}\label{sec:intro}
In the era of cyber-terrorism, the IT critical infrastructures have the crucial need for managing a huge amount of security information. In this context,  Internet Service Providers (ISPs) play a key role in the security data handling process also in accordance with European directives on the lawful interception and data retention. 
The Security Information and Event Management (SIEM) systems, represent an indispensable solution in protecting and monitoring the assets of a critical infrastructure. Such systems are able to collect, analyze and aggregate log data from various and heterogeneous nodes in order to discover security issues.
Today, the SIEM architectures are going beyond the classical security domains and find application also as policy violation control systems for mobile telecom operators.
SIEM market is rapidly growing as highlighted in a dedicated Gartner technical report \cite{nic-gartner} showing a comparison between some SIEM solutions furnished by just as many vendors.
The SIEM provides a single security point of view and it is typically conceived as a two-zone system: a Security Information Management (SIM) module designed for log management/reporting, and a Security Event Management (SEM) for incident management and real-time monitoring.
As before mentioned, a SIEM is able to gather and analyze security logs generated by different network devices (often called probes)  such as routers, firewalls, host and network intrusion detection systems, business applications and so on, in order to perform both inter and intra-layer security analysis. The former, is achieved by correlating logs belonging to different layers of the OSI model whereas the latter, is accomplished by correlating the only logs belonging to a specific layer of the OSI model.
The log correlation (performed by correlation engine, the core of the SIEM) is needed to reveal  events  as, for example, the presence of encrypted real-time traffic (as Skype) that in many cases is used to circumvent any kind of legal monitoring activities. 
In this work we illustrate a novel (and developed from the scratch) probe called \textit{ESkyPRO} (Enhanced Skype traffic Probe) able to reveal the encrypted Skype traffic by exploiting a machine learning based approach.
The paper is organized as follows: \textit{Section 2} presents some related works; \textit{Section 3} discusses the SIEM architecture and introduces the new enhanced probe called \textit{ESkyPRO}; in \textit{Section 4} we describe the interaction between \textit{ESkyPRO} and SIEM core, along with some implementation details. 
\textit{Section 5} provides details about adopted machine learning-based classification methodologies and sketches a performance analysis. \textit{Section 6} ends the paper by drawing some conclusions.

\subsection{Motivation}\label{sec:mot}
Network infrastructures are considered a strategic asset for telecommunication operators that, even for legal issues, have to comply with specific security policies concerning on one hand, the protection from potential cyber-attacks \cite{Matta2017} and, on the other, the traffic tracking and monitoring as imposed by regulatory measures.
One of these basic policies concerns in denying (or limiting) encrypted real-time communications between users that result in a lack of legal control for telecom operators.  
It is well known that some modern technologies based on Voice over IP (VoIP) concept as Skype, allow to elude any sort of tracking activity, by implementing ciphering mechanisms able to be exploited also by non-skilled users.
By starting from the analysis of statistical features characterizing an encrypted real-time traffic, and then focusing on the popular Skype protocol, we have designed and developed \textit{ESkyPRO}, an innovative probe able to interact with OSSIM \cite{ossim-siem}, an open-source SIEM solution, conveniently customized by the authors of the present work. According to the best knowledge of authors, no similar open-source solutions seem to be present in literature. For the sake of clarity, the authors want to highlight that, although the realized probe has been designed to identify Skype traffic, the general principles can be extended to all kinds of multimedia encrypted traffic being the statistical flow features (eg. packet lengths, inter-arrival times etc.) preserved.   

%Inspired by such a hot topic, this work is aimed at proposing an evolution of classic SIEM systems, currently the best IT solution to face the security of critical network infrastructures. 
%As it will be discussed later in details, SIEMs provide a real-time analysis of security alerts generated by network appliances but, to the best authors' knowledge, they lack ad-hoc functionalities for detecting encrypted real-time flows (e.g. Skype).

%More specifically, our probe is able to cooperate with other network appliances in order to allow the SIEM to unveil encrypted Skype sessions and generate security alarms. 

\section{Related Works}\label{sec:rw}
%Il RW dovrebbe essere sulle tecniche per rilevare il traffico skype crittografato..
%Trattandosi di critical infrastructure credo che parte dei related debba riguardare i SIEM. Parte invece potrebbe riguardare la detection del traffico crittografato in modo che nel seguito si va a verticalizzare sulle proprietà della sonda (appartenente al SIEM).

Diverse scientific works concern SIEM as a critical part of an information security management system. This section is devoted at presenting a non exhaustive \textit{excursus} about the more relevant literature concerning the aspects we deal with.
%Diverse scientific works concern SIEM as a critical part of an information security management system. 
In \cite{Shirley2012}, the authors face the issue of bot detection by considering a local-host alert correlation method and propose to include such a method in the SIEM capabilities.
From an architectural point view, it is interesting the security system proposed in \cite{Cruz2015} based on the integration and cooperation between the domains of the ISP core infrastructure and the home network.
In \cite{Fredj2015} instead, the problem of alert correlation has been faced by considering a graph-based attack; in particular, the carried performance analysis shows that the system is able to correlate a huge number of alerts into a dozens of attack graphs, allowing to extract several attack properties with a good precision. 
The approach used in \cite{GhasemiGol2015} instead, is based on the development of an intrusion-alert correlation system according to the information included in the raw alerts without using any predefined knowledge; at this aim, the authors define the concept of alert partial entropy to find the alert clusters having the same information.
From the perspective of ciphered traffic detection instead, the authors in \cite{DiMauro2015} propose a machine learning approach to reveal an encrypted WebRTC traffic.
In \cite{5071820} the authors exploit
an association analysis to discover abnormal relationships between data gathered by SIEM in order to detect malware threats.
A collaborative approach to improve management and analysis of generated alarms by a SIEM is instead discussed in \cite{6060795}. In particular, 
the authors propose a federated model where different domains share information about attacks detected by their own SIEM systems in order to improve the global 
security of the infrastructure.
In \cite{Myers:2009:TIT:1558607.1558670} the focus is on the detection of malicious attackers able to exploit some information 
of web servers in a given enterprise. The aim of the research is to examine various approaches for detecting insider threat activities using 
standard tools and a common event expression framework.
An interesting approach is followed in \cite{igor} where statistical 
models combined with specific policies to detect cross-cutting security breaches are adopted. 
In \cite{6657301}, some security issues of an e-Health infrastructure for management of Electronic Health Records (EHRs) is debated. In particular, the paper proposes an enhanced probe designed to analyze data provided at different layers of the OSI model and a secure storage system designed to ensure integrity and unforgeability of stored data even if some architectural components are compromised by faults or attacks.
In \cite{garofalo}, authors present a novel SIEM system that integrates a enhanced decision support system and a resilient event storage system. The novel SIEM is customized for a specific critical infrastructure, namely a hydroelectric dam. An attack model that affects various portions of the information technology infrastructure of the hydroelectric dam is used to demonstrate that the SIEM system can significantly improve the cyber security of the monitored dam infrastructure.
It's worth noting that today a huge variety of SIEM solutions (74 as stated in \cite{siem-mosaic}) freeware and commercial are available.
OSSIM and Prelude \cite{prelude-siem} represent the most widely adopted SIEM systems. The former, provides features as vulnerability and risk assessment, network and host intrusion detection with file integrity monitoring as well as the the possibility 
to write correlation rules in XML format.
The latter, uses a correlation engine based on Python language that allows to write more complex correlation rules but doesn't provide risk assessment capabilities.
In our paper, we consider the core of OSSIM infrastructure enriched with new functionalities offered by an innovative probe for encrypted Skype traffic detection 
%\textcolor{red}{
%that exploits supervised machine-learning concepts. Actually, no other similar solutions in the open source SIEM world have been found.
%}

\section{Security Information and Event Management Architecture}\label{sec:architecture}

The deluge of IT services offered by the ISPs, makes it necessary to design such infrastructures taking into account a certain number of security requirements.
In order to protect critical systems as a Service Provider (or a part of it) or a corporate infrastructure, it is possible to exploit the capabilities of a SIEM.
One of the main purposes a SIEM should accomplish, is to find potential correlations among logs generated at the network level (routers, firewalls, etc.) and logs generated by applications, in order to detect sophisticated security issues that could potentially pass unnoticed by individual security devices. Additionally, SIEMs can take actions and trigger incident response procedures upon detection of a security issue. Most currently available SIEM systems provide the same basic features, except specific ones proposed by vendors.
The role of a SIEM system begins to be critical as the enterprise networks start growing
due to the addition of new devices, applications, or employees, and the number of events generated also boosts.
However, operating large-scale SIEM systems
require a large budget. A typical management platform in fact might cost US \$80,000, and an archival database might cost US \$20,000 as reported in \cite{Bhatt2014}.

As previously clarified, we have implemented a customized version of OSSIM SIEM in order to detect encrypted Skype traffic.
\begin{figure}[t!]
	\centering
	\includegraphics[scale=0.35]{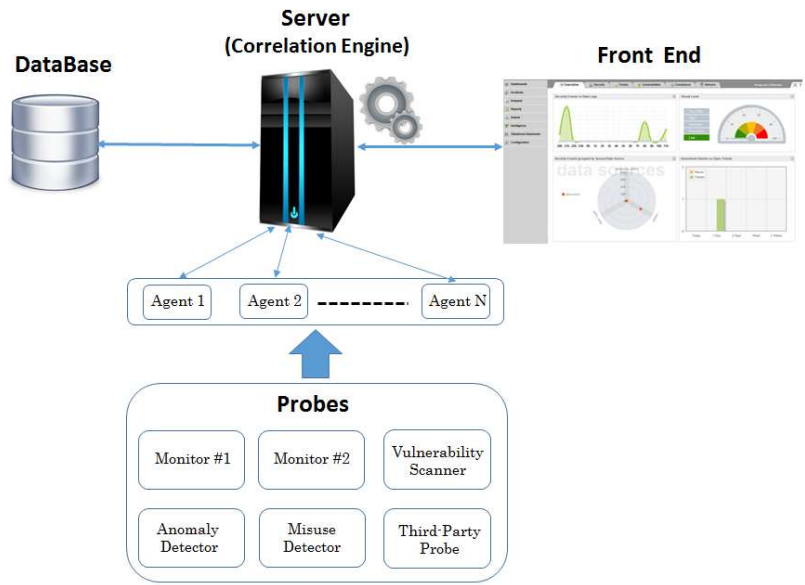}
	\caption{A classical architecture of a SIEM system.}
	\label{fig:architecture}
\end{figure}
In Figure \ref{fig:architecture} we depict a very common architecture of a SIEM, by  identifying the essential components:

\begin{itemize}
	\item \textit{Server}: represents the core component of the whole deployment, in charge of collecting and processing the logs coming from the external world on behalf of the correlation engine.
	\item \textit{DataBase}: stores all the data for the analysis and runtime configuration of the SIEM itself (basic modules configuration, taxonomies, asset tables and so forth).
	\item \textit{FrontEnd}:
	a console providing a user interface to the server. It furnishes to the security administrator a visual panel aimed both at controlling the single component configuration and at analyzing the security of the system under monitoring by means of dedicated dashboards.
	\item \textit{Probes}:
	a collection of sensors deployed within the monitored infrastructure. Typical examples of probes include perimeter defense systems (firewalls and intrusion prevention systems), host sensors (e.g. host IDSs) or security applications (web  firewalls and authentication systems).
	\item \textit{Agents}:
	the counterparts of probes embedded in the server, and able to convert heterogeneous logs generated by different probes, in logs with the same syntax and a specific semantic.
\end{itemize}

A probe can be deployed as a software or hardware element able to retrieve information from IT components such as routers, firewalls, web servers, anti-virus systems, intrusion detectors, in order to produce analyzable logs.
Typically, probes work in two modes: active and passive. In the active mode, the monitored IT component is not able to generate logs, so the probe needs to actively retrieve information by performing specific queries. In the passive mode, on the contrary, the monitored  component is able to generate and send logs to the probe so no ad-hoc queries are needed. Once logs are retrieved, each probe can perform preliminary security analyses by leveraging such information. In case of a security issue, an alert is generated by the probe itself and an informational log is sent to the corresponding  agent representing the entry point of the SIEM architecture. Three specific activities can be fulfilled by agents: normalization, aggregation and filtering. Normalization is performed by peculiar parsers and concerns a process to manipulate logs and alerts delivered by probes, in order to exhibit homogeneous data formats towards the server. Aggregation and filtering are exploited to reduce the logs volume aimed at simplifying the whole log analysis process.
The server has a module called correlation engine involved in the correlation rules processing; it includes a certain number of directives describing how to detect the most significant well-known security breaches. The purpose of the aforementioned engine is to check such rules in order to find suspicious relationships among logs and to raise alarms if those conditions are matched.
In the following subsections we describe two probes involved during the encrypted Skype flows revealing process. The first one is Snort \cite{snort}, a well-assessed Intrusion Detection System, whereas the second is \textit{ESkyPRO}, the novel and realized from the scratch probe. In \textit{Section 5}, some details about the cooperation between two probes will be provided.

\subsection{SNORT - Intrusion Detection System}\label{snort}
Snort is an open source intrusion detection system able to monitor and protect some strategic assets of an infrastructure \cite{rehman2003intrusion}.
The Snort architecture consists of the following modules: \textit{i)} packet decoder, \textit{ii)} preprocessors, \textit{iii)} detection engine, \textit{iv)} logging/alerting system and \textit{v)} output modules. The packet decoder, representing the entry point of Snort, captures data packets from different interfaces (e.g. Ethernet, PPP) and prepares them to be processed by detection engine. Preprocessors are dedicated plug-ins used by Snort to manage data packets in order to perform some preliminary security analyses as detecting anomalies within packets header. Another role in charge of a preprocessor is the packet fragmentation devoted at dividing a packet having an MTU (Maximum Transfer Unit) greater than 1500 bytes, in more sub-packets that need to be reassembled before applying any kind of rule.
Detection engine is a crucial and time-critical component of the whole Snort architecture designed to detect a potential intrusion activity after a rule match has occurred. A Snort rule includes some statements describing the patterns to analyze (type of protocol, header fields, payload information etc.) in a data packet; if packets match these rules some actions can be taken (e.g. alert, logging, dropping etc.).
Output modules at last, provide logs in a specific format (typically \emph{Syslog}) that will be forwarded to a remote server (the SIEM in our case) for further analyses.

\subsection{ESkyPRO: Enhanced Skype traffic Probe}\label{skypeDetector}

A very appealing probe realized on the basis of a previous work \cite{Longo2015} and embedded in the proposed SIEM infrastructure, has the main purpose of revealing in a statistical manner the Skype ciphered traffic. Skype is a VoIP application that allows to exploit some multimedia services \cite{Bonfiglio2009} as voice/video communication, chat and file transfer by encrypting the end-to-end flow \cite{Freire2008}. 

Even if the communication between users is established using a traditional end-to-end IP paradigm, Skype can also provide a routing mechanism towards a super node to facilitate the traversal of NATs and firewalls systems. A super node is an entity preserving the Skype overlaying network; in this scenario four types of peer-to-peer communications are allowed: client to client, client to super node, client to login server (for user credential management) or super node to super node. At the time of connection, Skype is able to choose a random port on the user host to send data traffic thus preventing the exploitation of filters using a port-based admission criterion. The unique piece of information usable for detection purposes is related to the login phase that can be divided in four steps \cite{Zhang2010}: 
\begin{enumerate}
	\item \textit{Scanning Super Nodes:} the Skype client sends a UDP datagram (with size that typically lies between 25 and 39 bytes) to super nodes with value of \textit{0x02} in the position 12 and it does not change with connections, packet times and version.
	\item \textit{Connecting to super nodes:} receiving the first response by super node, the Skype client tries to establish a TCP connection with the aforementioned node through a randomly selected port. The super node involved is referred to as “the servant super node” of the Skype client.
	\item \textit{Connecting to conn.skype.com:} after establishing the connection, the Skype client sometimes sends an HTTP request to 80 port of \textit{conn.skype.com} web server in order to get the latest version of Skype.
	\item \textit{Login on servers:} the login servers store the account information of users. The Skype client makes user authentication and obtains the buddy list during this step. The connection to the login server is, in some cases, relayed through a super node and therefore invisible.
\end{enumerate}

Except for this initial login phase, the whole Skype traffic is not in clear. In order to statistically characterize the traffic, some studies \cite{Alshammari2010} suggest to use a variety of intrinsic features of Skype flows, namely: packet length, inter-arrival time and number of packets in forward and backward directions.
By taking into account some of these features, \textit{ESkyPRO} has been designed according a supervised machine learning-based approach aimed at discriminating the \textit{Skype} traffic from the \textit{Normal} (namely not Skype) traffic. It is well known that in a supervised learning model two phases are typically considered \cite{Nguyen2008}:
\begin{itemize}
	\item \textit{Training:} the learning phase in charge of examining a set of labeled data (training dataset) and building a classification model; in this specific case, the Skype probe has been "trained" with separate Skype and Normal traces. 
	\item \textit{Testing:} the model built in the training phase is used to classify new unknown instances (Skype and Normal in our case).
\end{itemize}

\begin{figure}[t!]
	\centering
	\includegraphics[scale=0.28]{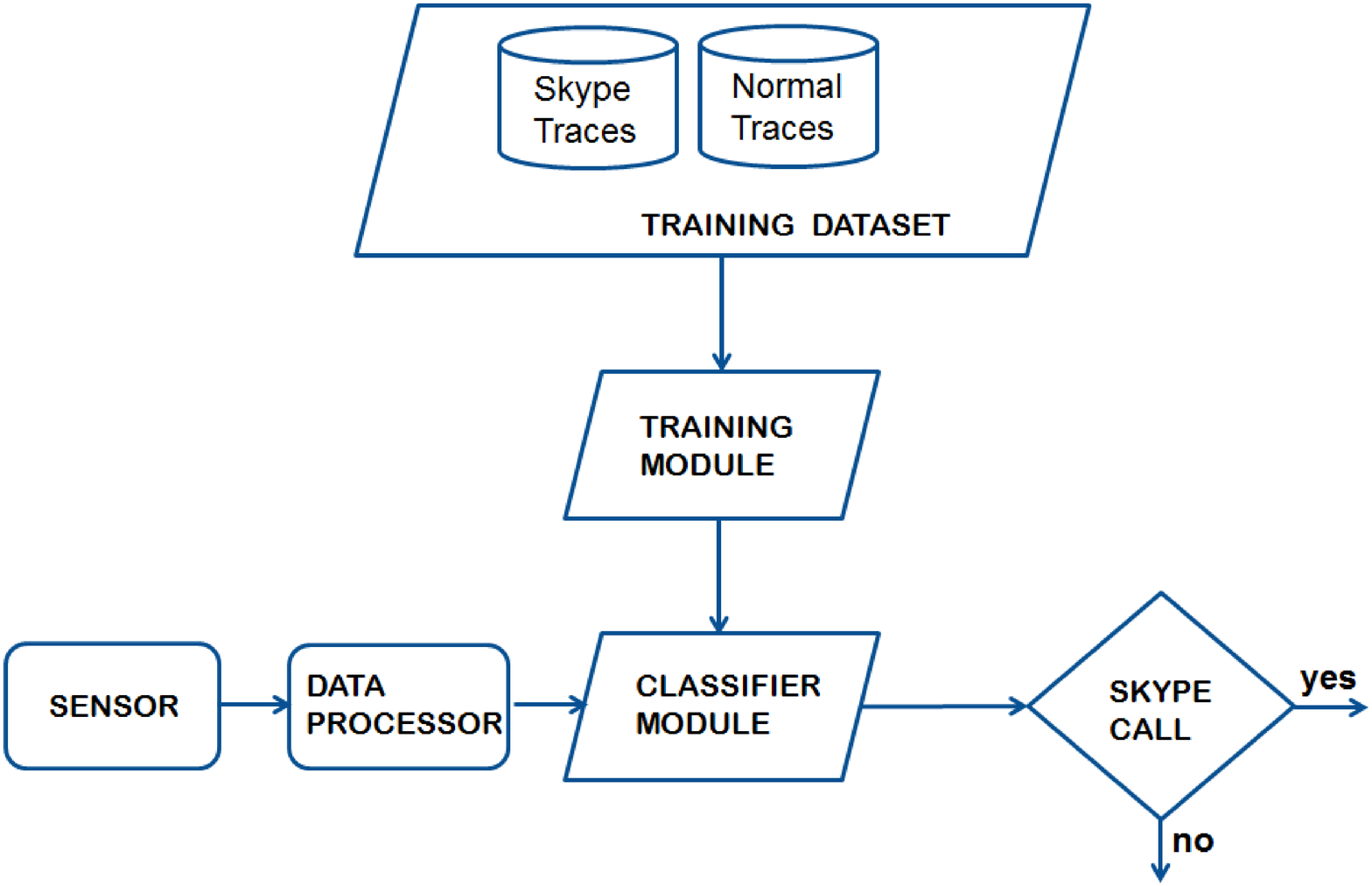}
	\caption{Logical architecture of \textit{ESkyPRO}.}
	\label{fig:probe_logic}
\end{figure}

The first step is to characterize the traffic flows on behalf of some attributes (or features), aimed at building an adequate training dataset. Such features, collected in a vector (say $\bf{F}$), are enumerated as follows:  
\begin{itemize}
	\item \textit{Proto: }Transport Level protocol type (TCP,UDP);
	\item \textit{AvgLgt:} average of packet lengths;
	\item \textit{StdLgt: }standard deviations of packet lengths;
	\item \textit{MinLgt, MaxLgt: }minimum and maximum packets lengths;
	\item \textit{AvgIAT:} average of inter-arrival times
	\item \textit{StdIAT: } standard deviation of inter-arrival times.
	\item \textit{MinIAT, MaxIAT: }minimum and maximum inter-arrival times;
\end{itemize}
From an architectural perspective, \textit{ESkyPRO} can be represented as in Figure \ref{fig:probe_logic} and includes the following modules:
\begin{itemize}
	\item A Sensor module (based on well known \textit{pcap} libraries) aimed at sensing the data traffic.
	\item A Data Processor able to normalize the data received by the sensor.
	\item A Training Dataset including pre-loaded Skype and Normal traces characterized in terms of defined attributes.
	\item Two modules (Training and Classifier) built on behalf of Weka APIs implementing the core of the classification algorithm and whose output is a boolean value \textit{Skype/Normal}.   
\end{itemize}
The implemented probe works in a quasi-real-time fashion. A sliding window mechanism with a tunable period (5 minutes for test purposes) allows to gather the unknown traffic coming from the sensor; such a traffic is then passed to the classifier module. The classification phase follows a two-step approach: first, some algorithms are applied by taking into account the training dataset embedded in the probe, and then a \textit{majority voting} procedure allows to combine in an efficient way the classifiers outcomes. Before considering in depth the classification strategies, let us discuss more in details the interaction between \textit{ESkyPRO} and the SIEM infrastructure.

\section{ESkyPRO and OSSIM integration}\label{sec:integration}
In this section we discuss about the integration of \textit{ESkyPRO} with the SIEM core. 
Basically, the communication between the probe and OSSIM core is guaranteed by a socket-based channel.
In particular, \textit{ESkyPRO} establishes a TCP connection towards OSSIM server on port 40001 and, after the handshake phase is completed, sends the following message to OSSIM server: 

\begingroup
\fontsize{9pt}{10pt}\selectfont
\begin{center} 
	\begin{boxedverbatim}
		Connect id="1" type="web" version="3.1.4" 
		hostname="ossim-server" tzone="1"
	\end{boxedverbatim}
\end{center} 
\endgroup
Once OSSIM receives such a message, it replies with:
\begingroup
\fontsize{9pt}{10pt}\selectfont
\begin{center} 
	\begin{boxedverbatim}
		Ok id=1
	\end{boxedverbatim}
\end{center} 
\endgroup
Thus, the new connection is established and OSSIM server is ready to communicate with \textit{ESkyPRO}. 
The next step concerns the adaptation of logs format generated by \textit{ESkyPRO} in compliance with logs managed by OSSIM system.
In particular, an OSSIM event message could contain the attributes described in Table \ref{table:OSSIM-attributes}.
\begin{table}[!t]
	\centering
	\small
	\begin{tabular}{ | p{1,6cm} | p{5cm} | }
		\hline
		\textbf{Attribute} & \textbf{Description} \\
		\hline
		plugin\_id & Probe identifier \\
		\hline
		plugin\_sid & Event type  \\
		\hline
		type & Type of the probe (detector or monitor) \\
		\hline
		date & Event generation date provided by the probe  \\
		\hline
		sensor & IP address of the probe involved in the event generation \\
		\hline
		interface & Name of the network interface (e.g. eth0) \\
		\hline
		priority & Event priority defines the importance of the event and it is used for the risk calculation \\
		\hline
		protocol & Communication protocol used (e.g. TCP or UDP) \\
		\hline
		src\_ip, src\_port & IP address and port number of host source that attempts a communication with a host destination \\
		\hline
		dst\_ip, & IP address and port number  \\
		dst\_port & of the host destination \\
		\hline
		log &	Original or raw log \\
		\hline
		userdata\_1 ... userdata\_9 & These fields are used to store user information useful to perform a complex security analysis \\
		\hline  
		\hline
	\end{tabular}
	\caption{Attributes contained in the OSSIM event message.}
	\label{table:OSSIM-attributes}
\end{table}
The fields \textit{plugin\_id} and \textit{plugin\_sid} concern respectively the identifier of the probe that sent the event message and the type of event itself and are both mandatory. The fields \textit{protocol}, \textit{src\_ip, src\_port} and \textit{dst\_ip, dst\_port} contain network information about the protocol (\textit{protocol}) used to attempt/establish a communication between a source host (\textit{src\_ip, src\_port}) and a destination host (\textit{dst\_ip, dst\_port}). The nine fields (\textit{userdata\_1...userdata\_9}) are used to insert additional information, namely, information about the application under monitor or user data, that can be exploited by the correlation engine to perform security analyses. 
Table \ref{table:skype-event} describes the attributes contained in a log message generated by \textit{ESkyPRO}.
\begin{table}[!t]
	\centering
	\small
	\begin{tabular}{ | p{1,6cm} | p{5cm} | }
		%\multicolumn{2}{c}{\textbf{Attributes contained within a OSSIM event}} \\
		\hline
		\textbf{Attribute} & \textbf{Description} \\
		\hline
		ipAddr & IP address of the host involved in a Skype session \\
		\hline
		timestamp & Timestamp indicates when skype session was performed \\
		\hline  
		\hline
	\end{tabular}
	\caption{Attributes contained in a log message generated by \textit{ESkyPRO} probe.}
	\label{table:skype-event}
\end{table}
Accordingly, a new agent has been created and deployed into OSSIM backend platform, aimed at mapping the logs generated by \textit{ESkyPRO} probe and OSSIM events. In particular, the new agent exploits the following regular expression:
\begingroup
\fontsize{10pt}{15pt}\selectfont
\begin{center} 
	\begin{boxedverbatim}
		regexp="ipAddr=(?P<userdata1>[^,]+), 
		.*timestamp=(?P<userdata2>[^,]+).*") 
	\end{boxedverbatim}
\end{center} 
\endgroup
Such a regular expression extracts the fields \textit{ipAddr} and \textit{timestamp} from each log generated by \textit{ESkyPRO} and maps them into fields \textit{userdata1} and \textit{userdata2} of the OSSIM event message as shown in Table \ref{table:OSSIM-attributes}. 
The main purpose of such an operation is to obtain normalized logs aimed at preparing the correlation engine embedded in the OSSIM platform, to manage various messages received by different deployed probes. 

\subsection{Probes  cooperation}

\begin{figure*}[ht]
	\centering
	\includegraphics[scale=0.12]{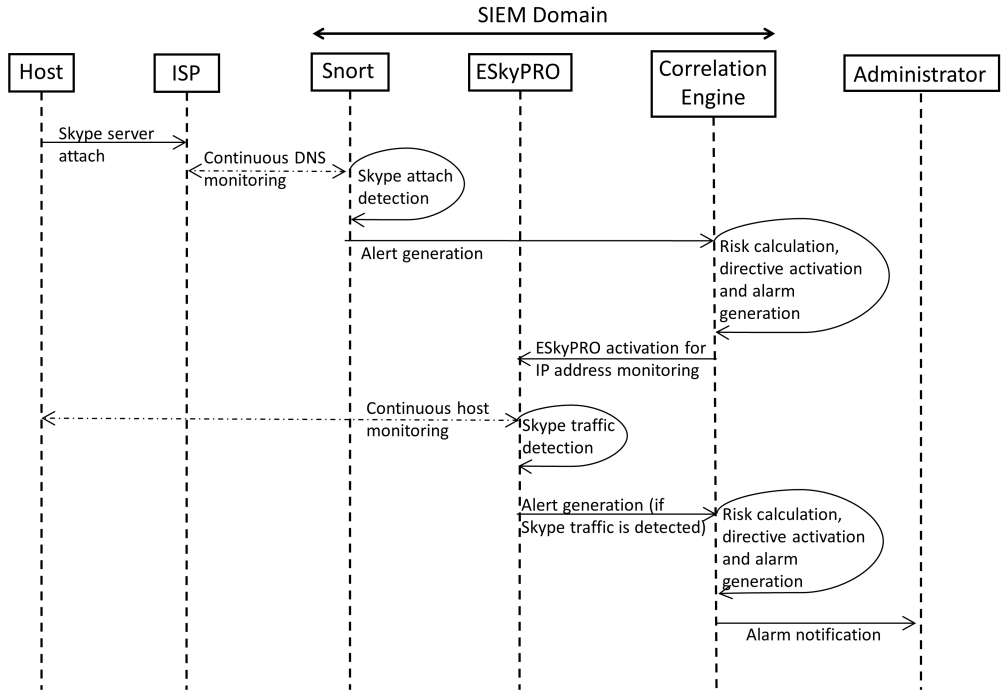}
	\caption{Sequence Diagram describing the Skype detection procedure.}
	\label{fig:sequence}
\end{figure*}

We are going to explain how the enhanced SIEM is able to detect an encrypted Skype traffic by correlating the logs produced by Snort and \textit{ESkyPRO} probes described respectively in Section \ref{snort} and in Section \ref{skypeDetector}.

As a representative scenario, we consider an ISP interested in detecting encrypted Skype traffic for two good reasons: \emph{i)} monitoring the usage of limited network resources; \emph{ii)} being compliant to some specific regulatory laws. As before mentioned, our testbed includes a customized implementation of OSSIM SIEM and two probes, namely Snort and \textit{ESkyPRO}. 
In order to better understand the steps performed by every component, a sequence diagram has been depicted in Figure \ref{fig:sequence}.
Snort probe is used to continuously monitor DNS (Domain Name System) queries performed by any host in order to detect if a login phase to the Skype server occurred that, we recall, it represents the unique information in clear travelling on the network.
It is worth noting that such an event (the login phase), doesn't entail an ongoing Skype call. 
The corresponding Snort rule to track the event is:

\begingroup
\fontsize{9pt}{10pt}\selectfont
\begin{center} 
	\begin{boxedverbatim}
		alert tcp any any -> any any (content: 
		"conn.skype.com"; msg: "skype login 
		attempt"; sid:1030; rev:1)
	\end{boxedverbatim}
\end{center} 
\endgroup
The rule instructs the Snort probe to monitor the connection from \textit{any} IP address and port towards \textit{any} IP destination and port over the \textit{TCP} protocol and contains the keyword \textit{conn.skype.com} denoting a common request automatically initiated by the client during the Skype server login phase. The rule also reports a \textit{signature identifier} (1030 in this case) and an informational message \textit{skype login attempt}.
In case of the rule is "fired", namely, a specific host initiating the login phase has been intercepted, the correlative \emph{Syslog} message including the intercepted host IP address and the timestamp of the performed request will be generated in the following form:

\begingroup
\fontsize{9pt}{10pt}\selectfont
\begin{center}
	\begin{boxedverbatim} 
		Syslog Snort log: {syslog} Mon Jan  
		16:11:50 CET 2017 INFO SnortSkypeAttach
		ipAddr=192.168.1.200#timestamp=16:11:45
	\end{boxedverbatim} 
\end{center} 
\endgroup

Such a message contains a login request to the Skype server by the host having the address \textit{192.168.1.200}.

This preliminary operation performed by Snort probe, lets \textit{ESkyPRO} able to monitor the only IP address involved in a potential Skype call with considerable saving of computational resources.

Once the SIEM correlation engine received the Syslog message from Snort, it calculates a numerical value called \textit{Risk}, a metric providing useful information about the overall security state of the infrastructure.
The risk value \textit{R} associated to an event, lies in the range [0-10] and it is calculated by the following formula: 
%\newpage

\begin{center}
	\small
	\textit{R} = $\displaystyle\frac{\textit{(Asset Value)} *  \textit{(Event Priority)} * \textit{(Event Reliability)}} {25}$
\end{center}

where: 
%\vspace{-11mm}
\begin{itemize}
	\item \textit{Asset Value} (0-5) takes into account the asset as hosts, host groups, networks and network groups;
	\item \textit{Event Priority} (0-5) defines the importance of the event itself;
	\item \textit{Event Reliability} (0-10) indicates the probability of a successful attack.
\end{itemize}
%\vspace{-0.85mm}
The parameters in the above formula can be tuned by experts of the considered domain.
When the Risk exceeds a defined threshold (say $R_{th}$) a security policy will be activated and some actions will be taken.
We recall that the main role of the correlation engine is to find relationships between logs in order to discover security breaches. If a security breach occurs, a new log having an updated value of priority and reliability is generated. 
Such an operation is specifically performed by correlation directives, representing one or more correlation rules defined as logical trees and represented by XML syntax. If a log causes a rule match, the next rule on the logical tree is evaluated up to the last one. 

The structure of the correlation directive used to detect a host performing a Skype server attach is shown below: 

\begingroup
\fontsize{9pt}{10pt}\selectfont
\begin{center}
	\begin{boxedverbatim}
		<directive id="501" name="Skype attach" 
		priority="5"> <rule type="detector" 
		name="attach discovery" reliability="3"
		occurrence="1" from="ANY" to="ANY" 
		port_form="ANY" port_to="ANY"
		plugin_id="4059" plugin_sid="1">
	\end{boxedverbatim}
\end{center}
\endgroup

%\begin{figure*}[t]
%	\centering
%	\includegraphics[scale=0.45]{tree}
%	\caption{A part of the decision tree generated by J48 algorithm}
%	\label{fig:j48}
%\end{figure*}

In the first lines, the identifier, the name of the directive and the numerical value of the directive priority have been reported;
the following lines indicate the rule type ("detector" indicates a plugin that actively sends information to the correlation engine) and the rule name; the next lines contain information about reliability value (3 in this case), occurrence value that indicates how many times the rule has been matched, the IPs/Ports under monitoring (ANY) and the identifiers of the probe that generated logs (plugin\_id = 4059, plugin\_sid = 1).

Once a directive has been matched, the correlation engine generates a log called \textit{alarm}. Assuming an Asset Value of 3, an Event Priority of 5 and Event Reliability of 3, the alarm Risk Value will be 1.8 and a security policy will be triggered to activate \textit{ESkyPRO} by passing the IP address of the host that attempted the Skype server attach.
At this time, \textit{ESkyPRO} starts to work waiting for a Skype session. 

\section{Classification logic in \textit{ESkyPRO}}

\textit{ESkyPRO} has been developed using a set of Java APIs offered by Weka \cite{weka}, an open source framework  developed at University of Waikato in New Zealand providing implementations of some machine learning algorithms. In order to give a certain degree of flexibility, we embed in our probe a set of $3$ classification algorithms representative of $3$ different classification philosophies whose outcomes, due to the complementarity of classifiers approach, are combined in order to provide a better performance \cite{Ho1994}. The three exemplary classifiers exploited in this works are: \textit{i)} J48 classifier, a decision tree method, \textit{ii)} the Logistic classifier, and \textit{iii)} the Bayesian Network Classifier. Each one of three classifiers have its \textit{pros} and \textit{cons} but, conveniently combined, they confer a certain robustness to the system. 
The mixing strategy exploited to obtain a combined result is the majority voting \cite{Kittler1998}, that appears as effective as it is simple to implement. 
In particular, this technique is useful when the individual classifiers give label outputs and when no huge volumes of data for classifier combinations are needed \cite{Kuncheva14},\cite{Lam97}. Roughly speaking, the majority voting only accounts for the most likely class provided by every single classifier by choosing the most frequent class label, where, aimed at avoiding ties, the number of classifiers involved in voting is usually odd. 
We want to highlight that \textit{ESkyPRO} is absolutely prone to embody other algorithms as well. 

\subsection{J48 classifier}

A very popular classification algorithm aimed at generating a so called decision tree based on a set of labeled training data is J48, a Weka's implementation of Quinlan \textit{C4.5} algorithm \cite{Quinlan1996}.
In some recent works \cite{Kalyani2012},\cite{Fernando2014}, J48 algorithm has been successfully exploited in classifying network traffic into ``normal" and ``abnormal" by considering a set of particular features aimed at revealing potential network anomalies with a reasonable accuracy. 

The resulting decision tree is a data structure consisting of decision nodes and leaves; a decision node specifies a test over a selected attribute whereas a leaf denotes a class value. The test on a continuous attribute \textit{A} has two possible outcomes: $A\leq t$ and $A > t$ where \textit{t} is a threshold determined at the node \cite{Ruggieri2002}. We recall that, the attributes considered in our scenario are those listed in Sect. \ref{skypeDetector}.
The J48 algorithm builds the decision tree with a \textit{divide and conquer} strategy that can be summarized as the following general pseudo-code \cite{Kotsiantis2006}:
\begin{enumerate}
	\item Check for base cases;
	\item For each attribute \textit{X} compute the Information Gain;
	\item Considered \textit{X-best} the highest normalized Information Gain attribute, create a decision node that splits on \textit{X-best};
	\item Recurse on the sub-lists obtained by splitting on \textit{X-best} and add those nodes as children of node.
\end{enumerate}
The Information Gain (IG of an attribute is defined by starting from the concept of \textit{entropy} (usually indicated by $H$), a measure of the impurity of a training examples collection \textit{T} that can be expressed as
\begin{equation}
H(T)=\sum_{j=1}^{c} -p_{j} log_{2}p_{j}, 
\label{entropy}
\end{equation}
where $c$ is the number of classes and $p_{j}$ is the proportion of \textit{T} belonging to \textit{j-th} class. 
Accordingly, the Information Gain of an attribute $X$ belonging to $T$ is defined as the reduction in entropy caused by partitioning $T$ according to such attribute, hence:

\begin{align}
	\label{eq:gain}
	IG(T,X)=H(T)-\sum_{values(X)}H(T_v) \frac{\vert T_v \vert}{\vert T \vert},
\end{align}
where \textit{i)} $T_v$ is the subset of $T$ for which attribute $X$ has value $v$ and \textit{ii)} $\vert T \vert$ and $\vert T_v \vert$ are cardinalities of $T$ and $T_v$ respectively.
The lower the entropy, the higher the value of an attribute and the corresponding Information Gain. 
A \textit{Pro} of J48 classifier concerns the aptitude to interpret the results but as a \textit{Con} it can exhibit a noteworthy complexity as the tree depth increases.

\subsection{Logistic Classifier}

Let us consider a two-class classification problem (as in our case) where $\omega_1$ and $\omega_2$ are the two admissible classes and where $\bf{X}$ is a vector.
In a probabilistic setting, the label of the class with highest posterior probability should be chosen, thus, the classifier would assign $x$ to $\omega_1$ if $p(\omega_1 \vert x) > p(\omega_2 \vert x)$, to $\omega_2$ otherwise. 
The logistic classifier \cite{Aggarwal2014}, \cite{Landwehr2005}, assumes a particular model for the class posterior probabilities, namely

\begin{align}
	\label{eq:logistic1}
	p(\omega_1 \vert X) =g(\beta^T X) = \frac{1}{1+ e^{(-\beta^T X)}},
\end{align} 
where
\begin{align}
	\label{eq:sigmoid}
	g(z)=\frac{1}{1+e^{-z}}
\end{align}
is called sigmoid function or logistic function.
As the sum of probabilities must equal $1$, we can also write

\begin{align}
	\label{eq:logistic2}
	p(\omega_2 \vert X) = 1 - g(\beta^T X) = \frac{e^{(-\beta^T X)}}{1+ e^{(-\beta^T X)}}.
\end{align}
Fitting a logistic regression model is equivalent to estimate the parameter vectors $\beta$ by exploiting the maximum likelihood criterion.
The aim is to maximize the data likelihood $L$ defined as

\begin{align}
	\label{eq:likelihood}
	L=\prod_{i=1}^{N}p(\omega_1 \vert x_i)^{n_1(x)}p(\omega_2 \vert x_i)^{n_2(x)},
\end{align}
where $n_k(x)$ is equal to $1$ if $x$ belongs to class $\omega_k$ and $0$ otherwise.
Since there is no closed form solution to maximizing $L$ with respect to $\beta$, in general a gradient descent method is used to solve the problem \cite{Snyman2005}.
In the classical Bias-Variance dilemma \cite{Geman1992}, the Logistic classifier exhibits low bias (\textit{Pro}) but suffers from high variance (\textit{Con}).

\subsection{Bayesian Network Classifier}

A Bayesian Network is a structure in which the attributes are graphically represented by nodes connected by directed edges having no cycles and forming a so-called Directed Acyclic Graph (DAG). The edges represent the relationships and dependencies between attributes. 
Let us assume a set $\Omega$ of classes $\omega_1,\dots,\omega_c$ and an attribute vector $x_1,\dots,x_n$.
According to the maximum a posteriori classification rule, we can write:

\begin{align}
	Class=\argmax_{\omega_j \in \Omega} p(\omega_j \vert x_1,\dots,x_n),
\end{align}
where the term $p(\omega_j \vert x_1,\dots,x_n)$ denotes the posterior probability of having the class instance $\omega_j$ given the evidence ($x_1,\dots,x_n$). Such a probability can be rewritten according the Bayes rule as

\begin{align}
	\label{eq:bn}
	p(\omega_j \vert x_1,\dots,x_n)=\frac{p(x_1,\dots,x_n)p(\omega_j)}{p(x_1,\dots,x_n)}.
\end{align}
Once built up a Bayesian Network, we need an algorithm for learning it. One of the most popular and used is $K2$ algorithm proposed in \cite{Cooper1992}.
By starting from a structure of attributes (nodes), the algorithm processes each node in turn and greedily considers adding edges from previously processed nodes to the current node. At each step, $K2$ adds the edge that maximizes the score of the network. When no further improvements are possible, the algorithm turns to the next node. The best resulting network is the one that maximizes the posterior probability.
The Bayesian Network classifier creates a network equivalent to the one used by Na\"ive Bayes algorithm when the maximum number of parents is equal to $1$. In this particular case, in fact, the attributes are considered as statistically independent, resulting in a loss of performances. For the purposes of this work, the Bayesian Network classifier works by considering a maximum level hierarchy of $3$, in order to account for dependencies among non-class variables. 
Bayesian Network classifiers allow to predict class labels when only partial information about input attributes are available (\textit{Pro}) but, on the contrary, require the number of parents as a tuning parameter, that can result in a performance loss if not adequately chosen (\textit{Con}).

%%%%%%%%%%%%%%%%%%%%%%%%%%%%%%%%%%%%%%%%%%%%%%%%%%%%%%%%%%%%%%%%%%

\subsection{Combining results of classifiers: the Majority Voting approach}
As previously stated, different classifiers are exploited to catch different behaviors of the system resulting in a final decision that takes into account various perspectives. Finally, all the outcomes are combined according to a majority voting scheme, whose technical details, are drawn in this section.
Let us consider a classification problem where the feature vector $\bf{F}$ is associated to one of $c$ possible classes ($\omega_1,\dots,\omega_c$). Let us suppose $R$ to be the number of classifiers and $x^{(i)}$ to be the measurement vector associated to the $i-th$ classifier. 
Bayesian Theory allows to state that $\bf{F}$ can be assigned to class $\omega_j$ by maximizing the a posteriori probability, namely
\begin{align}
	\label{eq:map}
	P(\omega_j \vert x^{(1)},\dots,x^{(R)})=\max_{k} P(\omega_k \vert x^{(1)},\dots,x^{(R)}).
\end{align}
Besides, according to the Bayes' Theorem, it is possible to write
\begin{align}
	\label{eq:bayes}
	P(\omega_k \vert x^{(1)},\dots,x^{(R)})= \frac{p(x^{(1)},\dots,x^{(R)} \vert \omega_k) P(\omega_k)}{p(x^{(1)},\dots,x^{(R)})},
\end{align}
where $p(x^{(1)},\dots,x^{(R)})$ represents the unconditional joint probability density that can be expressed as
\begin{align}
	\label{eq:condit}
	p(x^{(1)},\dots,x^{(R)})=\sum_{j=1}^{c}p(x^{(1)},\dots,x^{(R)} \vert \omega_j)P(\omega_j).
\end{align}
This latter corresponds to a representation in terms of conditional distributions.
By assuming that $x^{(i)}$ are statistically independent vectors (assumption verified in many cases, see e.g.  \cite{Kittler1997}), we can rewrite the joint probability distribution of measures as

\begin{align}
	\label{eq:stat_indip}
	p(x^{(1)},\dots,x^{(R)} \vert \omega_k)=\prod_{i=1}^{R}p(x^{(i)} \vert \omega_k).
\end{align}
Substituting from (\ref{eq:stat_indip}) and (\ref{eq:condit}) into (\ref{eq:bayes}) and, in turn, the resulting equation into (\ref{eq:map}), we obtain the following decision rule:

\begin{align}
	P(\omega_j) \prod_{I=1}^{R}p(x^{(i)}\vert \omega_j)=\max_{k=1}^{c}P(\omega_k)\prod_{i=1}^{R}p(x^{(i)}\vert \omega_k),
\end{align}
that, in terms of a posteriori probabilities can be expressed as:

\begin{align}
	\label{eq:mother}
	P^{(1-R)}(\omega_j)\prod_{i=1}^{R}P(\omega_j \vert x^{(i)}) \cr
	=\max_{k=1}^{c}P^{(1-R)}(\omega_k)\prod_{i=1}^{R}P(\omega_k \vert x^{(i)}).
\end{align}
Now, under the assumption that the a posteriori probabilities computed by classifiers would not differ hugely from the respective prior probabilities, we can write:

\begin{align}
	\label{eq:epsilon}
	P(\omega_k \vert x^{(i)}) = P(\omega_k)(1+\epsilon_k^{(i)}), 
\end{align}
with $\epsilon_k^{(i)} \ll1$. Substituting (\ref{eq:epsilon}) in (\ref{eq:mother}) leads to:

\begin{align}
	\label{eq:postepsilon}
	P^{(1-R)}(\omega_j)\prod_{i=1}^{R}P(\omega_j \vert x^{(i)})= \max_{k}P^{(1-R)}(\omega_k)(1+\epsilon_k^{(i)}).
\end{align}
By expanding the product and neglecting second (and higher) order terms, it is possible to approximate the R.H.S. of (\ref{eq:postepsilon}) and obtain:

\begin{align}
	\label{eq:semifinal}
	P(\omega_k)\prod_{i=1}^{R}(1+\epsilon_k^{(i)}) = P(\omega_k)+P(\omega_k)\sum_{i=1}^{R}\epsilon_k^{(i)}.
\end{align} 
The substitution of (\ref{eq:semifinal}) and (\ref{eq:epsilon}) in (\ref{eq:mother}) allows to obtain the following decision rule

\begin{align}
	\label{eq:final}
	(1-R)P(\omega_j)+\sum_{i=1}^{R}P(\omega_j \vert x^{(i)}) = \cr
	\max_{k=1}^{c}[ (1-R)P(\omega_k) + \sum_{i=1}^{R}P(\omega_k \vert x^{(i)}) ].
\end{align}
Let us now assume that all classes are a priori equiprobable and that a posteriori probabilities (say expert outputs) are hardened to produce binary values $\delta_k^{(i)}$ so that $\delta_k^{(i)}=1$ if $P(\omega_k \vert x^{(i)}) = \max_{j=1}^{m}P(\omega_j \vert x^{(i)})$ and zero otherwise, the majority will result in a decision for $\omega_j$ if

\begin{align}
	\label{eq:majority}
	\sum_{i=1}^{R}\delta_j^{(i)}=\max_{k=1}^{c}\sum_{i=1}^{R}\delta_k^{(i)}.
\end{align}
It is possible to observe that for each class $\omega_k$ the sum on the R.H.S. of (\ref{eq:majority}) counts the votes received for this hypothesis from individual classifiers. Therefore, the class receiving the largest number of votes, is chosen as the majority decision.

%%%%%%%%%%%%%%%%%%%%%%%%%%%%%%%%%%%%%%%%%%%%%%%%%%%%%%%%%%%%%%%

%the J48 algorithm and builds internally a decision tree where the nodes represent the flow attributes and the leaves represent the decision results (Skype/Normal). As an example, a sub-tree of the whole decision tree is shown in Figure \ref{fig:j48}.

%At each node (attribute) the algorithm makes a binary test and opts for one between two choices; at the end of tree a decision is taken.
%See the notation inside every leaf block: for example, Skype (63.0/1.0) means that 63 instances of “Skype” reached the leaf and only one is incorrectly classified. The tree creation procedure is often followed by a phase called pessimistic pruning, a top-down method allowing to simplify the decision tree by substituting a single leaf in place of a whole sub-tree when a specific irrelevance condition is met as suggested in \cite{kuhn2013}.

\subsection{System Performance}

It is worth recalling that the core of \textit{ESkyPRO} is represented by a decision system that \textit{i)} combines the outcomes of three classification algorithms by exploiting a majority voting rule and, \textit{ii)} reveals a potential ongoing Skype flow once a specific threshold value is exceeded.
Such a threshold, is strongly related to performance of majority voting algorithm, thus, this section is devoted at assessing a performance evaluation both for the three separate classifiers (J48, Logistic and Bayesian Network classifier) and for the Majority Voting classifier that acts as a fusion system.
All the algorithms have been tested on experimental data gathered from lab trials by exploiting a popular open-source data sniffer.
The collected traffic amounts to about 50 MBytes of Skype flows (30 Skype sessions) and about 50 MBytes of standard traffic with different protocols (HTTP, FTP, Streaming).
The resulting training set is obtained offline by extracting from the bulk of collected traffic some features (see Sect. \ref{skypeDetector}) that are then embodied in a vector containing $1292$ instances labeled with $Skype$ and $Normal$ tags. 
The performance results of various algorithms obtained by exploiting the aforementioned training set are summarized in Table \ref{table:tavoletta}. 
The True Positive Rate (TP Rate) and the False Positive Rate (FP Rate), indicate respectively the proportion of positive cases that were correctly classified and the proportion of negative cases that were incorrectly classified as positive, given a specific class (Skype or Normal).  
The MAE (Mean Absolute Error) is a measure of the average magnitude of the errors whereas the RMSE (Root Mean Squared Error) represents the squared and averaged difference between the expected and the observed samples. 
It is worth noting that the performances of various classifiers are enough balanced. For example, J48 exhibits the worst value of TP Rate for Skype class while outperforms other classifiers in terms of TP Rate for Normal class. Being a mix of different techniques, the Majority Voting classifier sometimes outperforms its contributor classifiers, but, at other times, underperforms them. 

\begin{figure}[t!]
	\centering
	\hspace*{-0.5cm} 
	\includegraphics[scale=0.65]{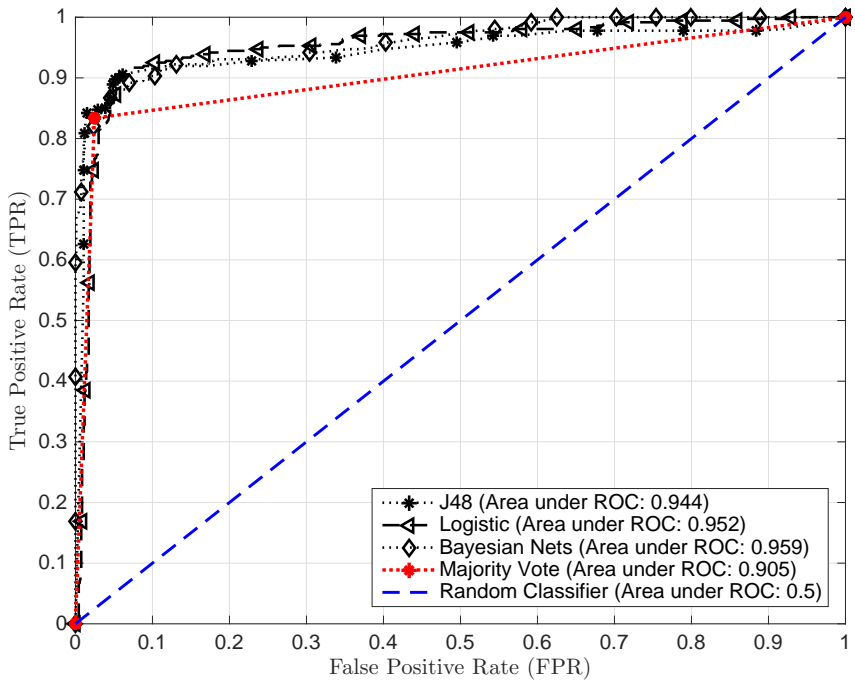}
	\caption{ROC Curves and Areas under ROC (AUC) for the three considered classifiers (J48, Logistic, Bayesian Networks) and the resulting majority vote-based classifier.}
	\label{fig:roc}
\end{figure}

\begin{table}[!ht]
	\centering
	\normalsize
	\begin{tabular}{ c | c | c | c | c }
		%\begin{tabular}{ m{1em} | m{1em} | m{1em} | m{1em} }
		\hline
		\multicolumn{5}{c}{\textbf{J48 Classifier}} \\
		\hline
		Class & TP Rate & FP Rate & MAE & RMSE   \\
		\hline
		Skype & 0.842 & 0.015 &  \multirow{2}{*} {0.0765} &  \multirow{2}{*} {0.2206}  \\
		Normal & 0.985 & 0.158 & \\ 
		\hline
	\end{tabular}
	\begin{tabular}{ c | c | c | c | c }
		\multicolumn{5}{c}{\textbf{Logistic Classifier}} \\
		\hline
		Class & TP Rate & FP Rate & MAE & RMSE \\
		\hline
		Skype & 0.856 & 0.047 & \multirow{2}{*} {0.1204} &  \multirow{2}{*} {0.243}  \\
		Normal & 0.953 & 0.144 &  \\
		\hline
	\end{tabular}
	\begin{tabular}{ c | c | c | c | c }
		\multicolumn{5}{c}{\textbf{Bayesian Network Classifier}} \\
		\hline
		Class & TP Rate & FP Rate & MAE & RMSE \\
		\hline
		Skype & 0.853 & 0.041 & \multirow{2}{*} {0.0739}  &  \multirow{2}{*} {0.2458}  \\
		Normal & 0.959 & 0.147 &  \\
		\hline
	\end{tabular}
	\begin{tabular}{ c | c | c | c | c }
		\multicolumn{5}{c}{\textbf{Majority Voting Classifier}} \\
		\hline
		Class & TP Rate & FP Rate & MAE & RMSE \\
		\hline
		Skype & 0.834 & 0.025 & \multirow{2}{*} {0.0642}  &  \multirow{2}{*} {0.2535}  \\
		Normal & 0.975 & 0.166 &  \\
		\hline
	\end{tabular}
	\caption{Classification results by applying various techniques. The outcomes concern \textit{i)} True Positive (TP) Rate, \textit{ii)} False Positive (FP) Rate, \textit{iii)} Mean Absolute Error (MAE), \textit{iv)} Root Mean Squared Error (RMSE). }
	\label{table:tavoletta}
\end{table}

%Even if \textit{J48} algorithm has been considered as an exemplary use case, the presented probe can also deal with other kinds of classification strategies. In particular, by exploiting our experimental setting, we have considered and compared two more classification algorithms as well: \textit{Na\"{i}ve Bayes}  \cite{Aggarwal2014} and \textit{Simple Cart} \cite{Breiman1984}.  	
%The \textit{Na\"{i}ve Bayes} classifier, based on the Bayes' rule, is well suited in case of high dimensionality of the input data. It relies on the class conditional independence assumption amongst attributes and takes a decision in accordance to the evidence provided by the training data. Thus, letting $a=(a_1,...,a_n)$ a vector of observed attributes and $C$ a class, such a classifier decides for $C_i$ class maximizing the a-posteriori probability:  	
%$P(C_i | a) = \frac {P(a | C_i) P(C_i)}{P(a)}$ 
%$P(C_i | a) = {P(a | C_i) P(C_i)} / {P(a)}$.
%The \textit{Simple Cart} algorithm, similarly to \textit{J48}, is based on a binary recursive partitioning procedure, being parent nodes splitted into exactly two children nodes. In place of Information Gain, the \textit{Simple Cart} classifier defines the \textit{Gini Index} as a selection metric, namely:
%$Gini(j)=1- \sum_{i=1}^{c} p^2_{i | j}$ where $p_{i | j}$ represents the probability that an instance in node $j$ belongs to class $i$. In case of all instances of training sample $S$ belong to the same class, then $p_{i | j}=1$ and the \textit{Gini Index} assumes $0$ value, corresponding to a node with no \textit{impurity}.

\begin{figure*}
	%	\begin{minipage}[b]{7cm}
	\centering
	%		\subcaptionbox{A frictionless ball in a semi-circle.}
	\begin{subfigure}[t]{0.4\textwidth}
		\centering
		\includegraphics[width=6cm]{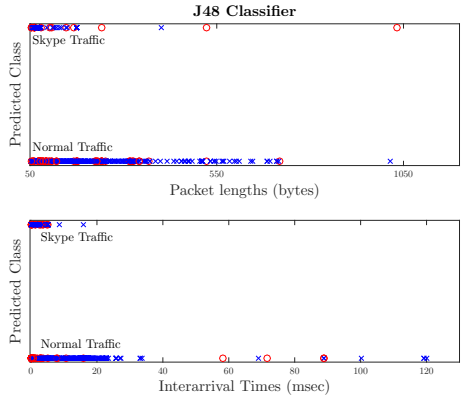}
		\caption{Errors in J48 Classifier}
	\end{subfigure}
	\vspace{4.5mm}
	\hspace{19mm}
	\begin{subfigure}[t]{0.4\textwidth}
		\centering
		\includegraphics[width=6cm]{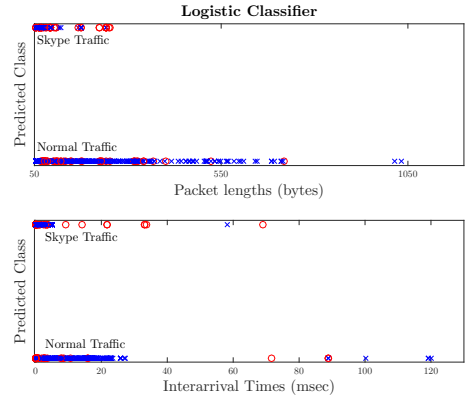}
		\caption{Errors in Logistic Classifier}
	\end{subfigure}
	\begin{subfigure}[t]{0.4\textwidth}
		\centering
		\includegraphics[width=6cm]{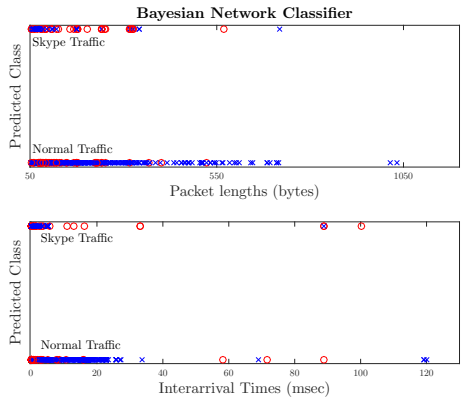}
		\caption{Errors in Bayesian Network Classifier}
	\end{subfigure}
	\hspace{19mm}
	\begin{subfigure}[t]{0.4\textwidth}
		\centering
		\includegraphics[width=6cm]{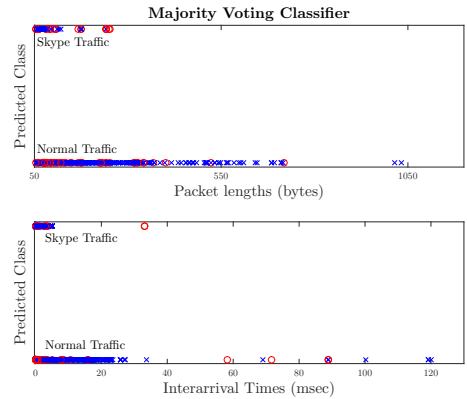}
		\caption{Errors in Majority Voting Classifier}
	\end{subfigure}
	\caption{Classification errors in terms of Packet lengths (bytes) and Interarrival times (msec) for: J48 Classifier (a), Logistic Classifier (b), Bayesian Network Classifier (c), Majority Voting (combining all the classifiers) Classifier (d). Blue cross symbols identify the correctly classified instances while red circle symbols identify misclassified instances.}
	\label{fig:errors}
\end{figure*}

Figure \ref{fig:roc} reports a set of curves stemming from the Receiver Operating Characteristic (ROC) analysis, aimed at investigating the relationship between false positive rate (FPR) measures and true positive rate (TPR) measures.
ROC analysis \cite{Egan1975} is exploited in many applications to address issues as determining a decision threshold minimizing the error rate, or characterizing regions where a classifier performs better than another one.
As a threshold value used to take a decision about a potential ongoing Skype traffic, \textit{ESkyPRO} considers the Area under ROC (AUC), a statistical indicator representing the probability that a randomly picked negative example have a smaller estimated probability of belonging to the positive class than a randomly picked positive example \cite{Hanley1982}. This indicator that provides a "summary" of a classifier performance, has been often adopted as a performance measure in medical trials, but recently has been proven to exhibit more precise outcomes than the accuracy in evaluating machine learning algorithms \cite{Huang2005}.  
In particular, when coping with imbalanced datasets, the AUC outperforms the accuracy being the latter not able to distinguish between the number of correctly classified examples of different classes.
The legend in Fig. \ref{fig:roc}, reports the AUC values for the considered classification algorithms. 
The larger AUC, the better the classifier's performance (AUC=1 means the classifier works perfectly).
A completely random classifier, namely a classifier whose decision is based on a coin flip, exhibits AUC=0.5 and it is depicted with a dashed blue line. As visible in Fig. \ref{fig:roc}, 
Majority Voting rule exhibits lower performances than other classifiers (AUC=0.905). This is due to the fact that all contributor classifiers (J48, Logistic, Bayesian Networks) are considered equally good and, if there are two of them giving the same incorrect label to a specific instance, the majority voting rule would favor the incorrect decisions \cite{Raykar2009}. This underperforming behavior is expected when combining classifiers implementing different philosophies but, on the contrary, it allows to build a robust threshold that allows \textit{ESkyPRO} to properly work by considering more "point of views". Such a threshold, in fact, encounters on one hand, the need of being not too strict (AUC too close to $1$) so as not to miss Skype flows that slightly differ from the expected behavior, and on the other hand, of being not too loose (AUC too close to $0.5$) in order to avoid the computational burden of the whole system, due to the false positives growth. Accordingly, we elect the AUC value of the Majority Voting classifier as a threshold ($AUC_{th}$) that, when overcame, allows \textit{ESkyPRO} to detect the presence of an ongoing Skype session. 
Aimed at offering a visual sketch of specific classification errors,  we choose two "pivot" features that typically cover a key role in characterizing multimedial flows as Skype, namely the packet lengths and the inter-arrival times. Figure \ref{fig:errors} reports this kind of analysis for all the different classifying strategies. Each plot reports on the y-axis the two predicted classes: Skype traffic and Normal traffic. A blue cross symbol indicates that a specific instance has been correctly classified, whereas, a red circle symbol denotes a misclassified instance. As a general trend, it is possible to highlight two facts. First, the points representing packet lengths of Skype flows tend to thicken around the x-axis origin whereas, the points related to packet lengths of Normal flows tend to move. This expected behavior is in line with a typical multimedia (e.g. Skype) session, characterized by forwarding small-sized packets, thus resulting more flexible in managing packet losses. Second, the points representing Skype inter-arrival times tend to be closer to x-axis origin as expected for a real-time flow, while, are spread over the x-axis in case of Normal traffic. As confirmed by values shown in Table \ref{table:tavoletta}, no dramatic differences between the classifiers exist in terms of performances.

\subsection{Detection procedure: the system at work} 

As previously clarified, the main outcome of classification engine embedded in \textit{ESkyPRO}, is a numerical value corresponding to the AUC value of Majority Voting classifier indicated by $AUC_{th}$. Such a value in fact, is used as a threshold value that, when overcame, forces the probe to advise the SIEM via \emph{Syslog} message, that a Skype session has been detected.
The Syslog message carries some informational parameters and turns up as follows:

\vspace{3mm}

%\vspace{-5mm}
\begingroup
\fontsize{9pt}{10pt}\selectfont
\begin{center}
	\begin{boxedverbatim}
		Syslog ESkyPRO log: {syslog} Mon Jan 30 
		19:25:23 CET 2017 INFO SkypeSession
		ipAddr=192.168.1.200#timestamp=19:25:30
	\end{boxedverbatim}
\end{center}
\endgroup
Such a message contains the IP address of the host involved in a Skype session, a time stamp and an informational message. 
Once the SIEM receives this message, the following correlation directive will be activated:

\vspace{3mm}

\begingroup
\fontsize{9pt}{10pt}\selectfont
\begin{center}
	\begin{boxedverbatim}
		<directive id="501" name="Skype session 
		detection" priority="5"> <rule type=
		"detector" name="Skype session detected" 
		reliability="5" occurrence="1" from="ANY" 
		to="ANY" port_form="ANY" port_to="ANY"
		plugin_id="4060" plugin_sid="1">
	\end{boxedverbatim}
\end{center}
\endgroup

The accomplishment of such a directive triggers a new alarm with an updated risk value of $R$=(3*5*5)/25=3 and a message is sent to the information center of SIEM (e.g. an administrator) that can decide to intervene and suggesting the best actions to perform.
It is worth noting that all numerical parameters used in this work have been tuned on the basis of the experience of technical experts but the system offers the possibility to tune such parameters if necessary. 

\begin{megaalgorithm}
	\caption{Skype traffic detection}
	\label{myalg}
	\KwIn{Packet Flows}
	%	\KwResult{how to write algorithm with \LaTeX2e }
	IDS initialization\;
	\While{flow contains  "conn.skype.com"  }{
		calculate Risk R\;
		\If{R $\geq$ $R_{th}$}{
			activate \textit{ESkyPRO} to evaluate AUC\;
			\If{AUC $\geq$ $AUC_{th}$} {Arise an alarm}
		}
		{continue the inspection\;}
	}
\end{megaalgorithm}

In order to better organize and summarize the overall procedures, an algorithmic representation of Skype detection processing performed by the SIEM with the \textit{ESkyPRO} probe embedded is shown in the Algorithmic Procedure \ref*{myalg}.

\vspace{-5mm}

\section{Conclusions}\label{sec:con}

In the era of the network data deluge, there is an ever  increasing need in managing security logs information coming from network devices. A Security Information and Event Management (SIEM) system exploits such data in order to provide a prevention plan aimed at securing data networks. Typically, a SIEM includes \textit{i)} a Database aimed at collecting the main logs coming from the network probes and \textit{ii)} a Correlation Engine in charge of discover potential relationships among the gathered logs.
In the present work, we introduce \textit{ESkyPRO}, a statistical network probe developed from the scratch and embedded on behalf of an agent counterpart in a customized version of OSSIM, an open source SIEM platform. 
An Intrusion Detection Probe based on a Snort implementation, acts as a preliminary trigger aimed at activating \textit{ESkyPRO} to monitor a particular IP address. 
Such a novel probe, includes a classification engine that takes advantage of machine learning techniques to reveal encrypted Skype traffic, traditionally impossible to detect by using classical intrusion detection analysis due to the absence of well recognizable patterns.
More specifically, such an engine implements three different classifiers (J48, Logistic and Bayesian Networks) whose outcomes are combined by leveraging a Majority Voting technique in order to strengthen the classification procedure.
The performance of Majority Voting classifier has been captured by means of Area Under ROC (AUC) indicator, whose value has been elected as a threshold parameter letting \textit{ESkyPRO} to recognize an ongoing Skype session.
Future works will be aimed to consider on the one hand, other network probes to involve in the encrypted traffic detection process and on the other hand, a wider class of algorithms to be included in \textit{ESkyPRO} with a special focus on the distributed ones.

\end{document}